# Impact of major hurricanes on electric energy production

**Julien Gargani** [1,2]

[1]Univ. Paris-Saclay, CNRS, Geops, Orsay, France

[2]Univ. Paris-Saclay, Centre d'Alembert, Orsay, France

julien.gargani@universite-paris-saclay.fr

**Abstract:**

After major hurricanes, electric production is significantly reduced owing to not only electric infrastructure destruction, but also the economic crisis associated with damage to private and public activities. The full restoration of the electric infrastructure is not always simultaneously performed with the full restoration of electric production. Here, we describe the electric production curves for the islands of Saint-Martin, Saint-Barthelemy, and Puerto Rico in the Caribbean, where two major hurricanes occurred in 2017. After the major hurricanes, the electric energy production was characterised by a slow recovery followed by a stable phase during several months, corresponding to approximately 75% of the initial electric production. A resilience time of several months (1 month < $t_s$ < 5 months) is necessary to attain the new electric energy production equilibrium $E_{is}$, which is lower than the initial electric energy production $E_i$. The reduction in electric energy consumption per capita is of 20–25%. In Saint-Martin, during the post-hurricane stable phase, the electric production was only approximately 60% of the initial electric energy production instead of 75%.

The reduction of ~15% in the electric production of Saint-Martin after Hurricane Irma could be attributed to the migration of approximately 8 000 inhabitants (approximately 23%) outside the island. This approach makes it possible to anticipate the production of electricity during several months after a major hurricane in the Caribbean islands.

1. Introduction

Major hurricanes can generate fatalities, damage infrastructures (roads, airports, harbours, electric networks), destroy dwellings, and devastate natural areas (mangroves, beaches, and trees) [Garnier and Desarthe, 2013]. Several studies have suggested that climatic change could increase hurricane intensity [Coumou and Rahmslorf, 2012; Hoarau et al., 2020]. The increase in the destructive power of hurricanes during the coming decades may have an impact on anthropic activities [Larsen et al., 2018]. Urbanisation of the coast has increased the vulnerability of French Caribbean islands to hurricanes during the last decades [Rey et al., 2020; Pasquon et al., 2020]. Consequently, the risks of destruction by hurricanes in the Caribbean are increasing. This study focuses on this area and particularly on Saint-Martin, Saint-Barthelemy, and Puerto Rico, which were affected by two major hurricanes, Irma and Maria, in 2017 [Duvat et al., 2019].

During major hurricanes, the destruction of buildings and infrastructure could be due to wind, heavy rain, landslides, or marine submersion (Yum et al., 2020). In this study, only hurricanes classified as type "5" on the Saffir–Simpson scale have been considered. During major hurricanes (classified as 5), wind speed of >250 km/h is expected. These winds cause enormous damage, especially in cases where buildings and infrastructure have not been adapted to this risk (small windows, resistant roof, use of concrete, etc.). Hurricane-proof buildings can reduce damage and increase the resilience [Prevatt et al., 2010]. The prevention of risks by adopting specific construction techniques and avoiding the location of sensitive infrastructure in coastal areas, evacuation of the population, rescue organisation, and post-crisis management can accelerate the restoration of the initial conditions. The adaptation [Walker et al., 2004] or lack of adaptation of societies to their environment can be observed using various long-term indicators such as demography or urbanisation in vulnerable areas [Gargani and Jouannic, 2015; Gargani, 2016; Jouannic et al., 2017].

The ability to return to a new equilibrium after a major devastation is of fundamental interest to societies. In this study, the considered duration is of several months before and after the hurricane occurred. The post-crisis period, spanning several months, is often considered as a strategic phase to reorganise territories

[Jouannic et al., 2016] and is defined as short-term adaptation. Long-term adaptation is beyond the scope of this study.

Resilience (that is, the capacity of the system to return to a new stable state) after a catastrophe is complex to observe [Bustillo et al., 2018]. Short-term resilience after a natural disaster can be monitored using several indicators such as night-time lights [Roman et al., 2019], health records (number of inhabitants affected by a specific disease), communication statistics (phone calls, SMS, etc.), water consumption [Gustin, 2018], and economic statistics (gross domestic product, sales of specific products). Economic statistics are generally monitored on a monthly basis and may not be sufficiently accurate. The observation of night-time lights using tele-detection methods for several months in areas where hurricanes occur provides spatial information about their resilience and allow to correlate spatial information with social arguments [Roman et al., 2019]. Nevertheless, many social and economic activities taking place in buildings, do not occur during the night (for example, schools) and cannot be detected by this approach. Because many indicators provide limited information, it may be interesting to introduce new indicators and analyse them taking into account other data. For example, monthly or yearly economic and/or demographic statistics (migration after the hurricane) can help interpret the evolution of other indicators with a daily or weekly acquisition period.

The destruction and reconstruction of the electric network after major disasters have been documented in several studies, making it possible to observe the challenges in rapidly processing a solution to basic needs [Kishore et al., 2018; Gustin, 2018;

Ramon et al., 2019; Der Sarkissian et al., 2021]. The destruction of the electric network reduces the electric energy production. Numerous deaths occur indirectly after natural disasters due to lengthy power outages and disruption of basic services [Kishore et al., 2018]. Long and recurrent electric outage favour morbidity increases (related to, for example, hospital dysfunction and reduced opportunities to communicate with doctors or social services) [Kishore et al., 2018]. The collapse of the electric network considerably reduces social and economic activities [Kishore et al., 2018]. To avoid external influence on electric energy production, such as import of energy that could modify local electric energy production, the case of three Caribbean islands was studied. Thus, the influence of exogenous factors that can increase the resilience of electric production was reduced.

The restoration of electric production is not influenced only by the reconstruction of the electric network. The reduction in social and economic activities could also generate a reduction in electric production several months after the electric network has been reconstructed.

Electric energy production can be an interesting indicator of the destruction and reconstruction of the electric network. Furthermore, it can indicate the resilience of social and economic activities. Indeed, during the past decades, reduction in the electric production in the Caribbean islands may be correlated with economic or social events, suggesting a causal link. For example, a reduction in the electric production and tourism in the Caribbean islands occurred (i) after the subprime mortgage crisis in 2008, (ii) after the September 11 terrorist attacks in 2001, and (iii) during the coronavirus epidemic (Covid-19) in 2020. Nevertheless, these three events did not impact the infrastructure in the Caribbean islands (no destruction of electric networks, hotels, houses, cars, or boats), and they differ from major hurricane impacts on electric production. The earthquake that occurred in Haïti in 2010 [Hayes et al., 2010] may have triggered the same type of electric production curve, but the data were not documented.

In this study, the capacity of territories to restore their activities after a major hurricane will be observed using electric production. The study focuses on the characterisation of electric energy production. A quantification of the parameters and processes involved in the restoration is conducted. The potential causes that can explain the delay in restoration of electric production are discussed.

The presence of an intermediate, stable electric energy production occurring a few months after major hurricanes has been observed in at least three cases (in Saint-Martin and Saint-Barthelemy after Hurricane Irma [EDF, 2018a and 2018b] and also in Puerto Rico after Hurricane Maria [US Energy Information Administration, 2019]) and is found to be intriguing. The origin and meaning of this pattern have never been explained. We question the possibility of this pattern being predicted and explained.

The aim of this study is to characterise the influence of major hurricanes on electric energy production from the moment of collapse until a new equilibrium was established, as observed in the electric energy production curves (Figure 1). To the best of our knowledge, no similar study has been conducted before. The evolution of electric production after a major hurricane in the case of three Caribbean islands (Saint-Martin, Saint-Barthelemy, and Puerto Rico) is described here to understand the main features of the post-catastrophe evolution and anticipate the consequences of electric production. This study proposes estimating the restoration rate of electric energy production after a major hurricane using a set of equations.

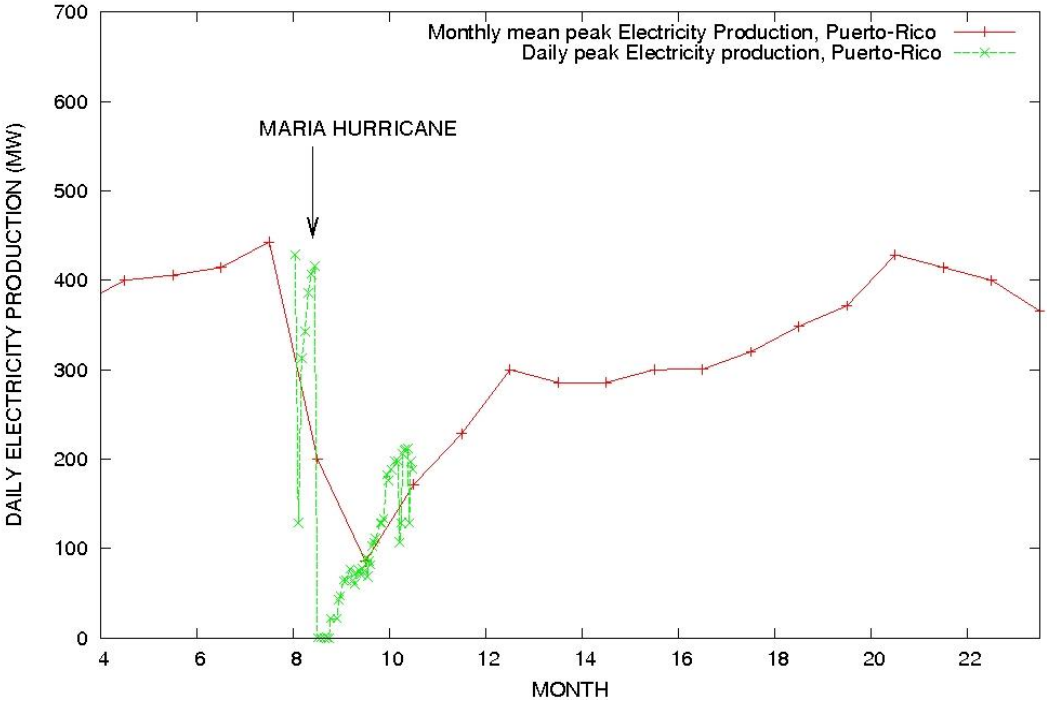

*Figure 1. Electric energy production of Puerto Rico before and after Hurricane Maria 2017: Daily electric production (green) and normalised monthly electric production (red). The daily electric production curve provides an accurate pattern just after the hurricane. The normalised monthly electric production makes it possible to observe the resilience of the electric system during a longer time. The monthly electric production has been divided by 30 to be compared with the daily electric production. Hurricane Irma is also visible in the daily electric production curve several days before Hurricane*

*Maria. During 2017, January ranges between 0 and 1 whereas September ranges between 8 and 9.*

**2. Context and Materials**

Hurricanes are classified on a scale proportional to the speed of their wind [Saffir, 1973; Simpson, 1974]. When the speed of the wind is limited, the destruction is generally limited, and the initial conditions are restored within a few days or weeks. For example, the restoration of the electric production in Puerto Rico after Hurricane Irma (September 6, 2017) was achieved in two weeks, considerably in a lesser duration than following Hurricane Maria (September 20, 2017) (Figure 1). Indeed, Hurricane Irma was not centred around Puerto Rico, but Hurricane Maria was. In contrast, Hurricane Maria was not centred around the islands of Saint-Martin and Saint-Barthelemy, but Hurricane Irma was. Both hurricanes were classified as 5 on the Saffir–Simpson scale.

This study focuses on two major hurricanes (Irma and Maria) on three different islands (Saint-Martin, Saint-Barthelemy, and Puerto Rico), where a reduction in electric production of more than 90% occurred. In the latter case, the infrastructure and social and economic activities were deeply impacted. There was significant damage to the electric network and destruction of private buildings [Jouannic et al., 2020; Roman et al., 2019]. The selected hurricanes were classified as 5 on the Saffir–Simpson scale. Hurricane Irma affected Saint-Martin and Saint-Barthelemy on 5 and 6 September

2017, whereas Hurricane Maria affected Puerto Rico one 20 September 2017. The winds were associated with marine submersion and heavy rain.

There are approximately 35 000 inhabitants in Saint-Martin, 10 000 inhabitants in Saint-Barthelemy [Servans et al., 2017], and 3.1 million inhabitants in Puerto Rico. Saint-Martin (France) and Sint-Maarten (The Netherlands) are located on the same island. In this study, we focus only on the French side of the island because the electric production is independent, and data were available only for the French side. Tourism is the main economic activity in Saint-Martin and Saint-Barthelemy. Puerto Rico has a more diversified economic activity.

There is a mean daily electric production of 27.5 MW in Saint-Martin [EDF, 2018a] and 16.5 MW in Saint-Barthelemy [EDF, 2018b]. The electric production capacity of Saint-Martin is $E_{max} = 56.6$ MW, but only 27.5 MW is produced [EDF, 2018a]. In Saint-Martin, the ratio $\tau_0$ between the electric production capacity $E_{max}$ and the effective electric production $E$ (i.e., electric energy really produced) was 0.48 before the hurricane and 0.30 after. The electric production capacity $E_{max}$ of Saint-Barthelemy is $E_{max} = 34.2$ MW, but only 16.5 MW is produced [EDF, 2018b]. In Saint-Barthelemy, the ratio $\tau_0$ between the electric production capacity $E_{max}$ and the effective electric production $E$ was 0.48 before the hurricane and 0.35 after. The electric production capacity $E_{max}$ of Puerto Rico is approximately 6 000 MW [US EIA, 2020], but only 400 MW were produced before Hurricane Maria. In Puerto Rico, the ratio $\tau_0$ between the electric production capacity $E_{max}$ and the effective electric production $E$ was 0.06 before the hurricane and 0.05 after.

The data on electric production are available for Saint-Martin (French side) and Saint-Barthelemy from Electricité De France [EDF, 2018a and 2018b]. Daily electric production is presented here, but the precision needed for the modelling can be less precise, especially during stable phases. Hurricane Irma had a significant impact on the electric production of Saint-Martin and Saint-Barthelemy [EDF, 2018a and 2018b]. The electric network was mainly aerial before the hurricane and most of it was destroyed by the wind. The electric production of Saint-Martin was also impacted by marine submersion because the production unit was located close to the sea [EDF, 2018a]. However, no major destruction of the production units occurred in Saint-Martin and Saint-Barthelemy.

The data from the U.S. Energy Information Administration were used to study the influence of Hurricane Maria on electric production in Puerto Rico [US EIA, 2020]. Hurricane Irma also influenced the electric production of Puerto Rico over a few days, but with a moderate impact in comparison with that of Hurricane Maria. In Puerto Rico, the daily electric production [Puerto Rico Data Lab, 2020] is superimposed with the normalised monthly peak of electricity production to observe it over a longer period of time (Figure 1). Some of the electric production units were out of order in Puerto Rico after Hurricane Maria [Roman et al., 2019]. The electric production considered here takes into account only that generated by electric companies. Individual production of electricity by small electric generators or individual solar panels was not taken into account.

## 3. Quantified description

A modelling approach was conducted to investigate the electric production dynamics after the occurrence of major hurricanes. First, a theoretical model was developed. Then, several sensitivity tests were performed to better understand the efficiency of the model and the role of the parameters. The model intends to estimate the electric production at various times based on several parameters.

The model assumes a competition between the accumulation and consumption of electric energy. More precisely, the rate of the electric energy production $V^a$ (in MWh/h) is counterbalanced by the rate of electric energy consumption $V_0^b$ (in MWh/h). The electric system is considered at equilibrium (i.e., the ratio $V^a / V_0^b$ is constant) until the collapse generated by the hurricane generates an abrupt decrease in the electric energy production. Starting from a low value, the electric energy production is constrained by the rate of electric energy production. The rate of the electric energy production $V^a$ changed after the hurricane occurrence and reached a new value, lower than the previous one.

The electric production curve shows a stable plateau after several months. This specific pattern must be reproduced as well as the trajectory of the electric production from the main blackout until the stable plateau. These patterns can be described by a mathematical function.

The electric production can be described by :

$$E = [\tau_0 + a \times ln\ (V/V_0) + b \times ln(V_0 \times \theta / L)] \times E_{max}, \qquad (1)$$

where $E_{max}$ is the maximum electric power capacity of the system (in MW), $V^a$ is the rate of the electric energy production, $V_0^b$ is the rate of the electric energy consumption, and $\tau_0$ is the ratio between the effective electric energy production and the maximum electric energy production capacity of the system $E_{effective} / E_{max}$. $L$ is the characteristic electric power (in MWh) of the system. The smaller the value of $L$ is, the better the resilience of the electric production. Parameters $a$ and $b$ are constants with $0 > a > b$. The amplitude of the combined parameter $a + b$ influences the loss of energy from the initial stable condition until the new equilibrium, after which the electric production collapses. $t$ represents time, and $\theta$ is a characteristic time and a state variable (a state variable can be used to describe the state of a dynamical system; see for example Scholz, 1998) of the system defined by :

$$\Delta\theta / dt = 1 - V \times \theta / L \qquad (2)$$

First, at the equilibrium $\Delta\theta / dt = 0$, and thus $1 - V \times \theta / L = 0$

$$\Leftrightarrow V = L / \theta$$

Including it in Equation (1), the electric production at the equilibrium becomes

$$E_{effective} = [\tau_0 + a \times ln(V/V_0) + b \times ln(V_0/V)] \times E_{max}$$

$$\Leftrightarrow E_{effective}/E_{max} = \tau_0 + (a-b) \times ln(V/V_0)$$

$$\Leftrightarrow E_{effective}/E_{max} - \tau_0 = 0 = (a-b) \times ln(V/V_0), \text{ by definition of } \tau_0$$

$$\Leftrightarrow ln(V/V_0) = 0$$

$$\Leftrightarrow V = V_0$$

$$\Leftrightarrow V^a = V_0^a = (V_0^b)^{a/b}, \qquad (3)$$

Consequently, at equilibrium, the competition between the rate of electric production $V^a$ and the rate of electric energy consumption $V_0^b$ is described by a power law (Equation 3). Post-cyclonic-electric-energy production is a non-linear process.

Second, when the hurricane occurs and the electric system collapses, there is a discrepancy between the electric energy production and electric energy consumption. The condition necessary to restore the electric system is that the rate of the electric production $V^a$ be greater than the rate of electric energy consumption $V_0^b$. According to Equation (1), the following can be written:

$$E/E_{max} - \tau_0 = a \times ln(V/V_0) + b \times ln(V_0 \times \theta /L ) \qquad (4)$$

During the collapse, the electric energy is not at equilibrium. There is a relaxation of the system. After the collapse, during the restoration of the electric energy production, an intermediate phase of equilibrium is observed (for example, in Figure 2). During this phase, the ratio between the electric production and the maximum capacity of the electric energy production $E / E_{max}$ is constant and equal to $\tau_0$.

Consequently, from Equation (4) we obtain

$$a \times ln(V/V_0) + b \times ln(V_0 \times \theta /L ) = 0.$$

We can deduce that $exp[a \times ln (V/V_0)] = exp[-b \times ln(V_0 \times \theta /L)]$.

Thus, during the restoration of the electric energy production, the rate of the electric energy production is equivalent to

$$V^a = ( V_0^b )^{a/b - 1} \times [ \theta /L ]^{-b}, \qquad (5)$$

Let us describe some basic consequences of Equation (5). First, the rate of the electric energy production $V^a$ and the rate of electric energy consumption $V_0^b$ are generally not equal. Second, if $\theta / L > 1$, then the rate of the electric energy production $V^a$ is greater than the rate of the electric energy consumption $V_0^b$ because a / b < 1. Under these conditions, the electric system will have no risk of collapse during the restoration of the system. Conversely, if $0 < \theta / L < 1$, the rate of electric energy consumption $V_0^b$ will be greater than the rate of electric energy production $V^a$, and a blackout can be expected. Third, in our model, the values of $V^a$ and $V_0^b$ are relative to each other, as given in Equation (5). In this approach, only the relative values between $V^a$ and $V_0^b$ play a role and not the absolute values. From Equation (3), it is possible to estimate the value of the rate of the electric energy consumption $V_0^b$ at equilibrium. In the transient state, it is necessary to use Equation (5).

A numerical code was developed in Fortran to solve Equations (1) and (2) using a finite-difference method. The electric production is relaxed by a change in the rate of electric production. For example, in Saint-Martin, the rate of the electric production $V^a$ before the hurricane was 1.65 MWh/h, whereas the rate of electric production $V^a$ after the hurricane was 1.08 MWh/h. An abrupt change in the rate of the electric production is necessary to simulate the observed trend. The spatial dimensions of the restoration will not be investigated.

## 4. Results

The daily electric production of Saint-Martin, Saint-Barthelemy, and Puerto Rico has been quantified before and after the occurrence of a major hurricane during a period of time of 8 months.

*4.1. Saint-Martin electric production*

In the island of Saint-Martin, the electric network was critically damaged by Hurricane Irma on 5 September 2017. The simulated daily electric production shows a resilience time of 2 months after the collapse of the electric network (Figure 2). Two different stable daily electric productions can be observed, before ($E_i$ = 27.5 MW) and after ($E_{is}$ = 17 MW) the electric blackout. The second stable phase of daily electric production (intermediate stable electric production, $E_{is}$) is lower than the first one, suggesting a deep modification in the activities and organisation of Saint-Martin.

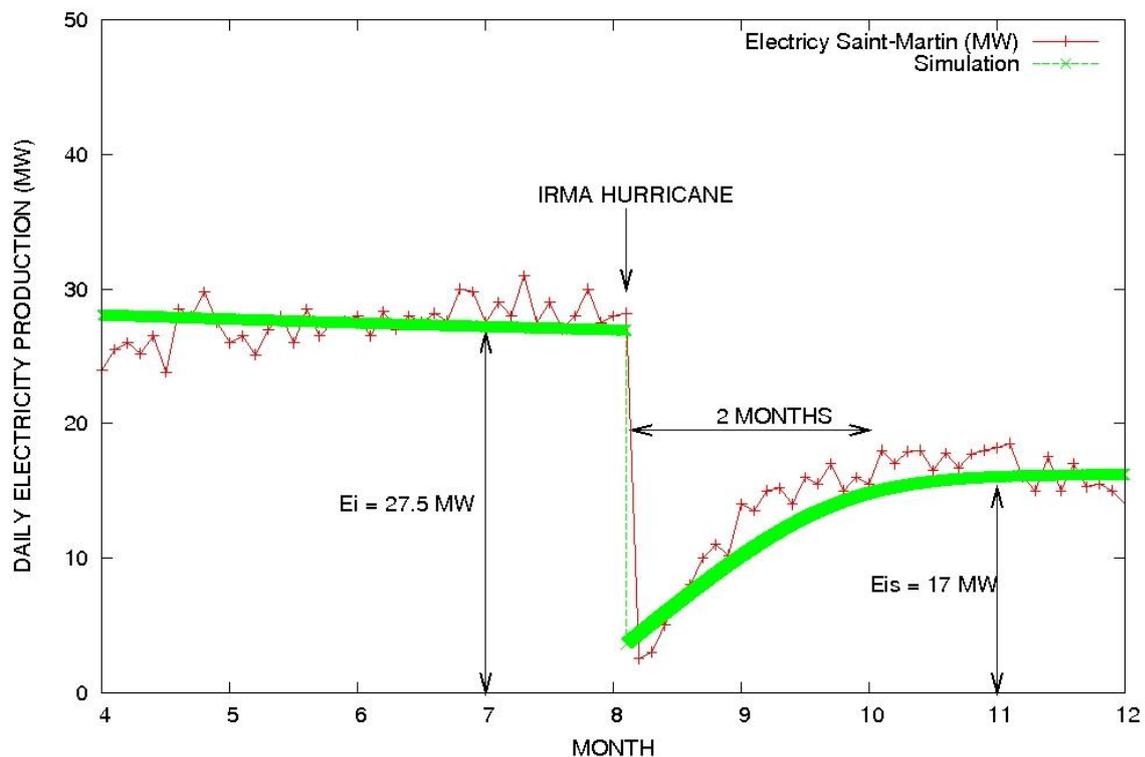

*Figure 2. Influence of Hurricane Irma on the daily electric production of Saint-Martin. Two stable states can be observed for the daily electric production. The first*

*one is approximately $E_i$ = 27.5 MW and the second one is approximately $E_{is}$ = 17 MW. The second intermediate stable plateau can be observed 2 months after the blackout due to the collapse of the electric network. The progressive restoration of the electric network is simulated with $E_{max}$ = 56.6 MW, $\tau_0$ = 0.30, V = 0.25 (MWh/h)$^{1/a}$, $V_0$ = 0.025 (MWh/h)$^{1/b}$, a = −0.054, b = −0.063, and L= 0.09 MWh. At t = 0, $\theta_t$ = 1 h.*

For Saint-Martin, the simulation of the daily electric production was performed using Equation (1), considering that the electric production capacity is $E_{max}$ = 56.6 MW. A value of effective electric production ratio $\tau_0$ of 0.30 was considered. In the modelling of Saint-Martin, the rate of electric energy production was $V^a$ = 1.65 MWh/h, the rate of electric consumption was $V_0^b$ = 1.08 MWh/h, and the electric power was L= 0.09 MWh. Using these parameters, the model could fit the electric production for 8 months before and after the impact of Hurricane Irma (Figure 2).

*4.2. Saint-Barthelemy electric production*

Even in the island of Saint-Barthelemy, located approximately 10–15 km east of Saint-Martin, the electric network was damaged by Hurricane Irma. The general trend of daily electric production shows the same pattern in Saint-Barthelemy than that in Saint-Martin. Two stable periods can be observed in the daily electric production of Saint-Barthelemy, the first before the blackout and the second one month after the main blackout (Figure 3). The resilience time is approximately 1 month from the blackout until the new stable equilibrium is reached. The first initial stable daily electric production $E_i$ is approximately 16.5 MW. The second intermediate stable equilibrium of the daily electric production $E_{is}$ is of 12 MW for at least 3 months.

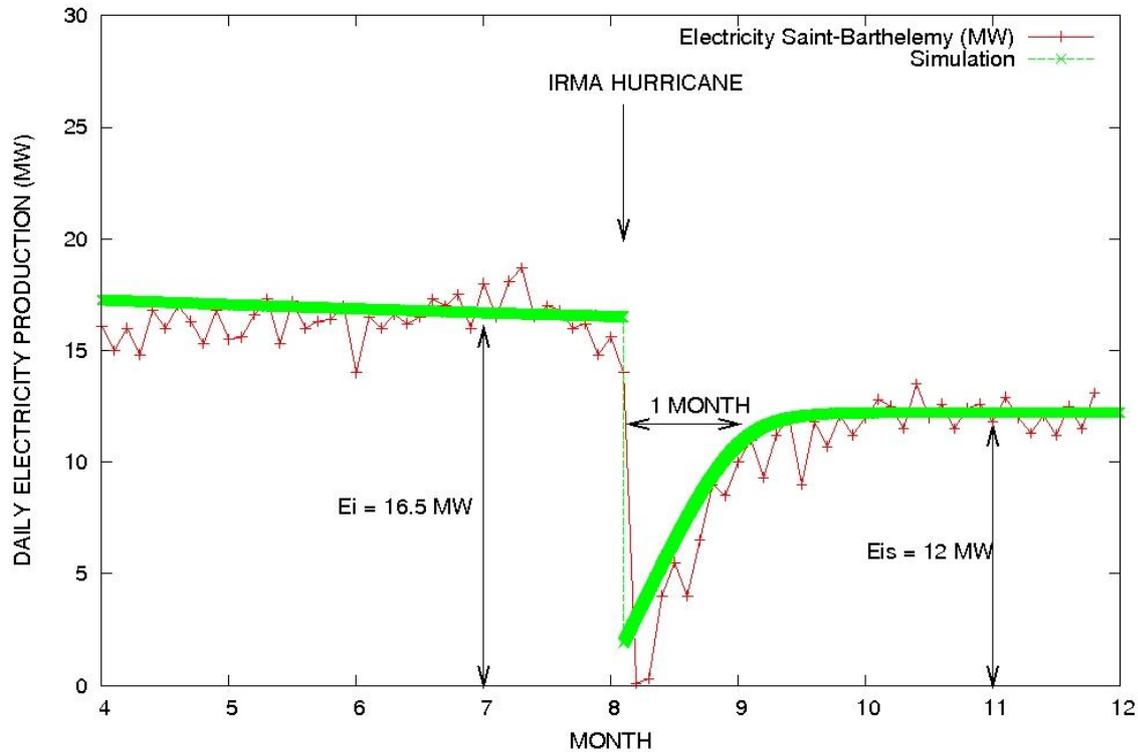

*Figure 3.* Influence of Hurricane Irma on the daily electric production in Saint-Barthelemy. Two stable states can be observed for the daily electric production. The first one is approximately $E_i$ = 16.5 MW and the second one is approximately $E_{is}$ = 12 MW. The second intermediate stable plateau can be observed 2 months after the blackout due to the collapse of the electric network. The progressive resilience of the electric network is simulated with $E_{max}$ = 34,2 MW, $\tau_0$ = 0.35, $V$ = 0.25 $(MWh/h)^{1/a}$, $V_0$ = 0.025 $(MWh/h)^{1/b}$, $\theta$ = 1 h, $a$ = −0.054, $b$ = −0.07, and $L$ = 0.05 MWh.

*4.3. Modelling Puerto Rico electric production*

In Puerto Rico, the transient state between Hurricane Maria and the new equilibrium extended over 5 months. Equations (1) and (2) provide the estimates of electric production before and after the hurricane at each time step (Figure 4). Before Hurricane Maria, the daily electric production was approximately 400 MW, whereas the daily electric production after the hurricane was 300 MW in the stable state.

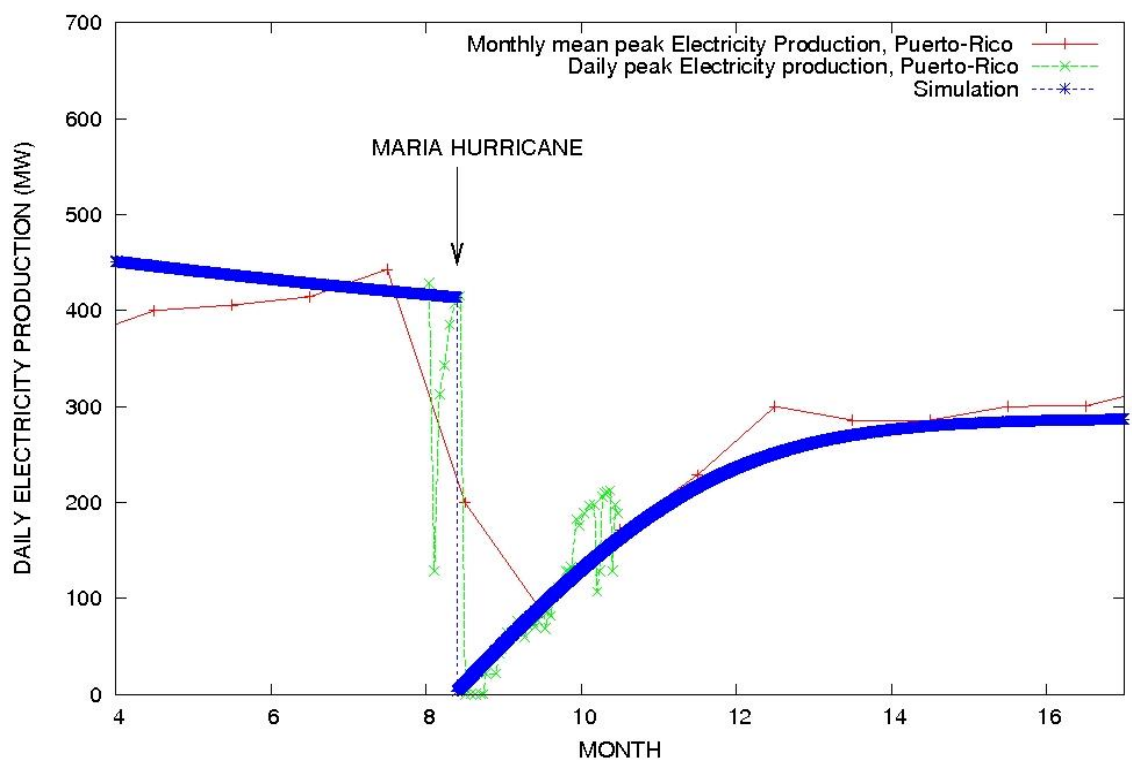

***Figure 4.*** *Influence of Hurricane Maria on the daily electric production of Puerto Rico. Two stables states can be observed for the daily electric production. The first one is approximately $E_i$ = 400 MW and the second one is approximately $E_{is}$ = 300 MW. The second intermediate stable plateau can be observed 5 months after the collapse of the*

*electric network due to Hurricane Maria. The progressive resilience of the electric network is modelled with $E_{max}$ = 6 000 MW, $\tau_0$ = 0.05, $V$ = 0.5 $(MWh/h)^{1/a}$, $V_0$ = 0.055 $(MWh/h)^{1/b}$, $\theta_t$ = 1 h, a = −0.008, b = −0.019, and L= 0.6 MWh. The monthly electric production was divided by 30 to obtain the equivalent daily electric production data.*

*4.4. Characteristic of electric energy production after major hurricane*

The pattern of the curves of electric energy production can be characterised by a few parameters such as:

- Resilience time $t_r$

- Electric resilience ratio $ER_R = E_i / E_{is}$

- Electric production restoration rate $E_{is} / t_r$,

- Offset resilience of the electric production $E_i - E_{is}$

These parameters allow us to characterise the restoration of stable electric energy production. Here, the resilience time $t_r$ is not the time necessary to restore the initial state, but the time necessary to reach a new intermediate equilibrium. In this study, resilience time was reached when the electric production became stable at $E_{is}$. The resilience time $t_r$ of Saint-Barthelemy was smaller than that of Saint-Martin. The resilience time $t_r$ of Saint-Martin was smaller than that of Puerto Rico (Table 1).

The intermediate equilibrium of electric production has been characterised by several indicators. In Saint-Martin, the electric resilience ratio was $ER_R = E_{is} / E_i = 0.62$

(Table 1). The higher the $ER_R$ ratio is, the more the system has restored the initial electric capacity. In Saint-Barthelemy, $E_{is}$ = 12 MW and $E_i$ = 16.5 MW. The electric resilience ratio was $ER_R = E_{is} / E_i$ = 0.73. In Puerto Rico, $E_{is}$ = 300 MW and $E_i$ = 400 MW. The electric resilience ratio was $ER_R = E_{is} / E_i$ = 0.75. The resilience ratio $ER_R$ of Puerto Rico and Saint-Barthelemy was higher than that of Saint-Martin (Table 1).

The electric production restoration rate is estimated using the ratio between the intermediate electric production ($E_{is}$) and resilience time ($t_r$). This rate represents the velocity of the system to reach a new equilibrium. Puerto Rico and Saint-Barthelemy have a higher rate of restoration of electric energy production than Saint-Martin (Table 1). The electric production restoration rate is not strictly proportional to the electric power capacity (i.e., the initial electric production $E_i$) of the territory, as shown by the comparison between Saint-Barthelemy and Saint-Martin (Table 1). Saint-Martin has a lower electric production restoration rate, whereas the electric power capacity of Saint-Martin is higher than that of Saint-Barthelemy.

Table 1. Main characteristics of electric production resilience after a major hurricane.

| Caribbean Island | Resilience Time $t_r$ (month) | Initial Electric Production $E_i$ (MW) | Intermediate Stable Electric Production $E_{is}$ (MW) | Electric Resilience Ratio $E_i/E_{is}$ | Electric Production Restoration Rate $E_{is}/t_r$ (MW/month) | Offset Resilience of Electric Production $E_i-E_{is}$ (MW) |
|---|---|---|---|---|---|---|
| Saint-Barth. | 1 | 16.5 | 12 | 0.73 | 12 | 4.5 |
| Saint-Martin | 2 | 27.5 | 17 | 0.62 | 8.5 | 10.5 |
| Puerto Rico | 5 | 400 | 300 | 0.75 | 60 | 100 |

However, the offset of the electric energy production from the initial state $E_i$–$E_{is}$ was higher for Puerto Rico and Saint-Martin than for Saint-Barthelemy (Table 1). This parameter is related to the initial electric power capacity (i.e., initial electric energy production, $E_i$). In the three cases studied, the resilience time, $t_r$, is proportional to the system size (initial power capacity) but is not correlated to the electric resilience ratio $E_i / E_{is}$. However, not all the parameters investigated are proportional to the system size. For example, the electric resilience ratio is not related to the system size. The electric resilience ratio is higher in Saint-Barthelemy and Puerto Rico than in Saint-Martin, whereas the initial electric energy production $E_i$ is higher in Saint-Martin than in Saint-Barthelemy.

The intermediate equilibrium of the electric production is significantly lower than that before the major hurricane occurrence in the three Caribbean islands studied. This reduction ranges between 25% and 38%. The decrease in electric production (i.e., the offset resilience of electric production $E_i$–$E_{is}$) is proportional to the initial electric production $E_i$ (Table 1).

*4.5. Influence of a, b, L, V, and $V_0$*

The parameters *a*, *b*, *L*, *V*, and $V_0$ of Equations (1) and (2) influence the electric energy production. Simulations show that the electric power *L* has a strong influence on the electric production (Figure 5). The higher the electric power *L* is, the higher the offset between the two stables "plateaus" of electric production $E_i$ and $E_{is}$.

Furthermore, the higher the electric power $L$ is, the longer the resilience time $t_r$ (Figure 5). A value of effective electric power $L$ = 0.05 MWh was used to simulate the electric production in Saint-Barthelemy, of 0.09 MWh in Saint-Martin, and of 0.6 MWh in Puerto Rico (Table 2). The higher the system (i.e., initial electric production), the higher the electric power $L$ (see Tables 1 and 2). Even if the effective electric power $L$ is not strictly linearly proportional to the initial system capacity (or to the number of inhabitants), a link may exist that it could be interesting to investigate in further studies with more data (Tables 1 and 2).

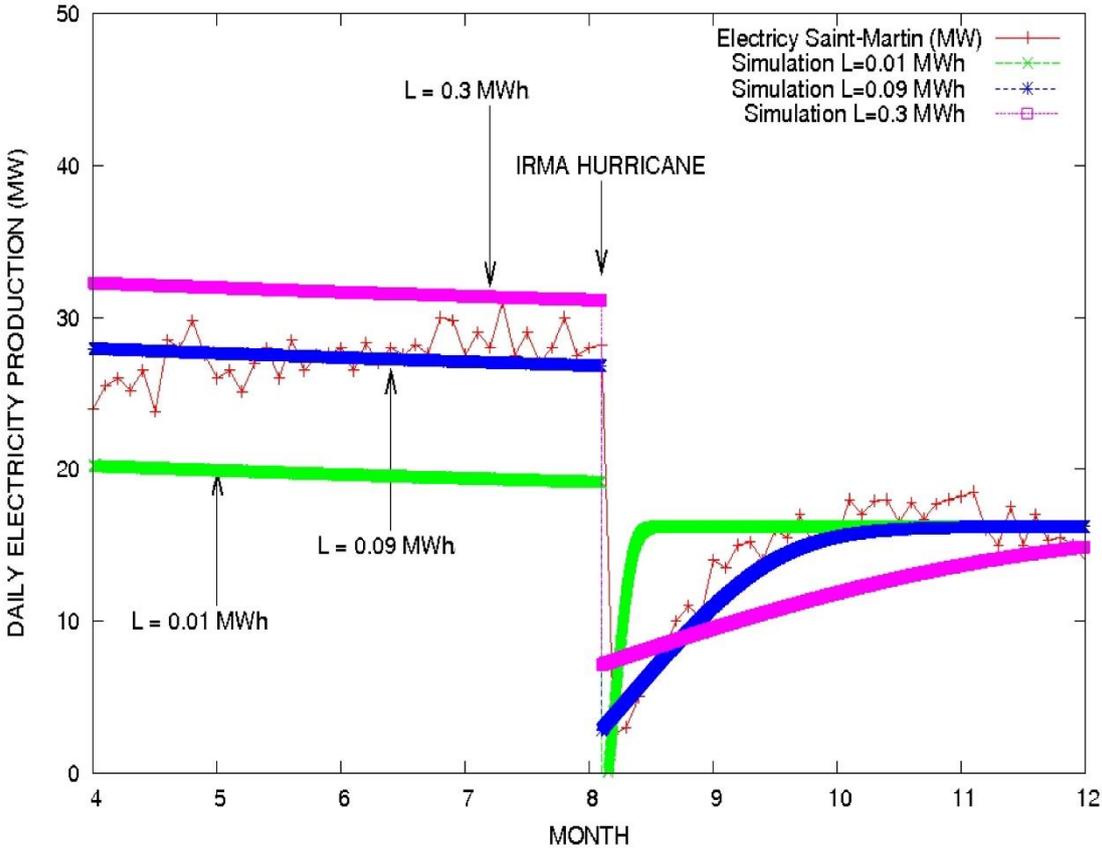

*Figure 5.* Influence of electric power L on electric energy production E. Other parameters used in the modelling: $E_{max}$ = 56.6 MW, $\tau_0$ = 0.30, V = 0.25 (MWh/h)$^{1/a}$, $V_0$ = 0.025 (MWh/h)$^{1/b}$, $\theta$ = 1 h, a = −0.054, b = −0.063.

The dimensionless constants $a$ and $b$ also play a significant role in the result of Equation (1). The constants $a$ and $b$ are negatives. The smaller the value of $|b|$ is, the higher the offset between the two stable values of the electric production $E_i$ and $E_{is}$ (Figure 6A, Table 1). In this study, $|b|$ is equal to 0.070 in Saint-Barthelemy (Figure 3), 0.063 in Saint-Martin (Figure 2), and 0.019 in Puerto Rico (Figure 4). The smaller the system size (i.e., initial energy distribution) is, the higher is $|b|$. There is an effect of $|b|$ on the increase in the electric production rate after the collapse. The higher $|b|$ is, the higher will be the increase in electric production after the hurricane. The higher $|a|$ is, the higher will be the offset between the two stable values of the electric production $E_i$ and $E_{is}$ (Figure 6B).

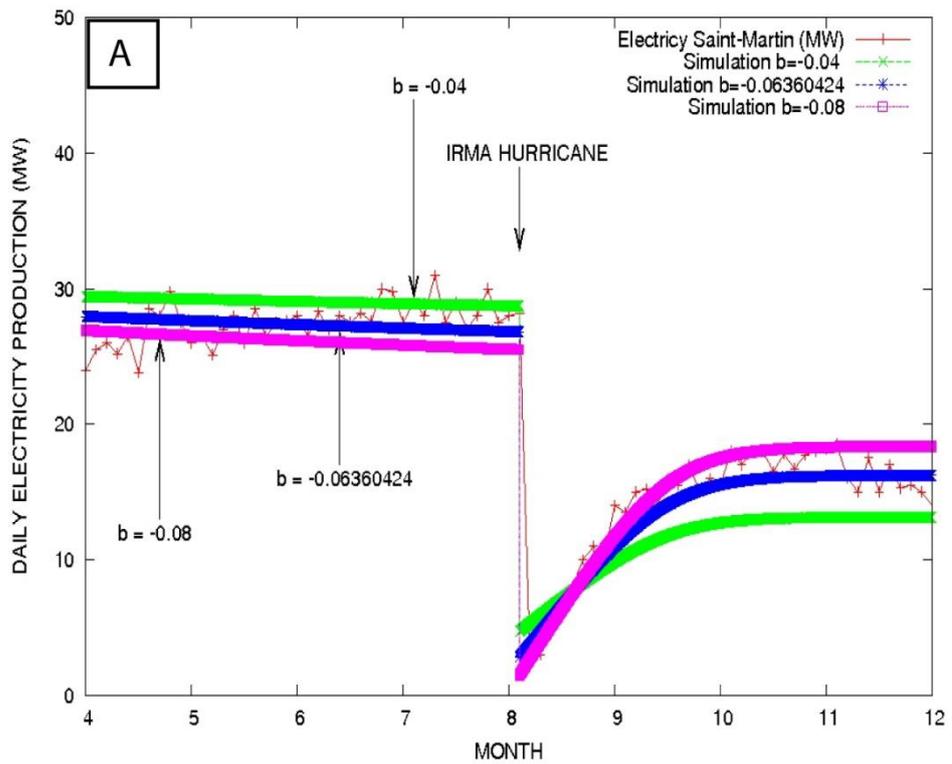

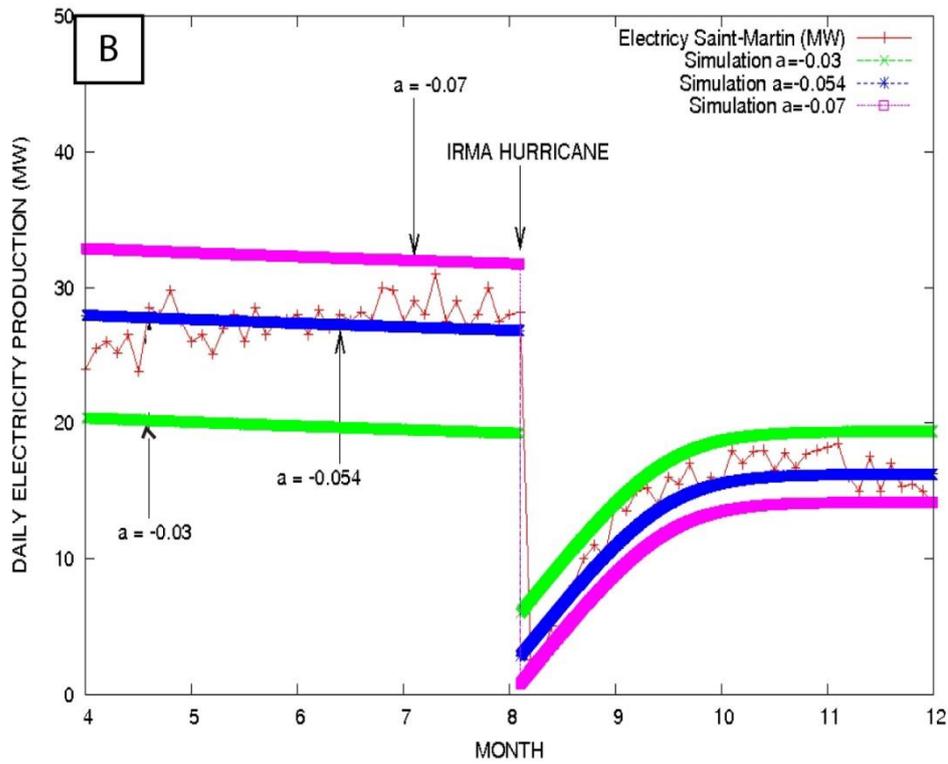

*Figure 6.* Influence of the adimensional constants b and a on the electric energy production. (A) Influence of the adimensional constant b on the electric production, a

= −0.054, (B) Influence of the adimensional constant b on the electric production, b = −0.063. The other parameters used in the modelling are $E_{max}$ = 56.6 MW, $\tau_0$ = 0.30, V = 0.25 (MWh/h)$^{1/a}$, $V_0$ = 0.025 (MWh/h)$^{1/b}$, $\theta_t$ = 1 h, and L= 0.09 MWh.

The higher the parameter V is, the faster the electric production reaches the intermediate stable electric production $E_{is}$ (Figure 7A). The resilience time of the system increases when V increases. In the calculation, V = 0.25 (MWh/h)$^{1/a}$. However, the absolute value of V is less important in the model than the ratio between V and $V_0$. There is an analytic relation between these two parameters at equilibrium, or when $\tau_0$ = E / $E_{max}$ is constant. $V_0$ is always smaller than V in the three cases studied, and the ratio V / $V_0$ ranges from 9.1 to 10 (Table 2). The variation in $V_0$ generates a translation of the electric production. The higher the value of $V_0$ is, the higher will be the electric production E (Figure 7B). The influence of this parameter can apparently be counterintuitive. However, this can be interpreted considering that a bigger country with a higher electric production E has a higher rate of electricity consumption as well.

Table 2. Main characteristics of the territories and parameters used

| Caribbean island | Inhabitants | Gross Domestic Product (Euro/capita) | Characteristic Electric Production L (MWh) | Plateau Combined Parameter a+b | Electric Prod./Consum. Rate Ratio V/$V_0$ | Electric Production Ratio (effect./max.) $\tau_0$ |
|---|---|---|---|---|---|---|
| Saint-Barth. | 10 000 | 40 000 | 0.05 | −0.124 | 10 | 0.35 |
| Saint-Martin | 35 000 | 16 600 | 0.09 | −0.117 | 10 | 0.3 |
| Puerto Rico | 3 100 000 | 27 000[1] | 0.6 | −0.027 | 9.1 | 0.05 |

[1] International Monetary Fund (dollars converted to euros; https://www.imf.org)

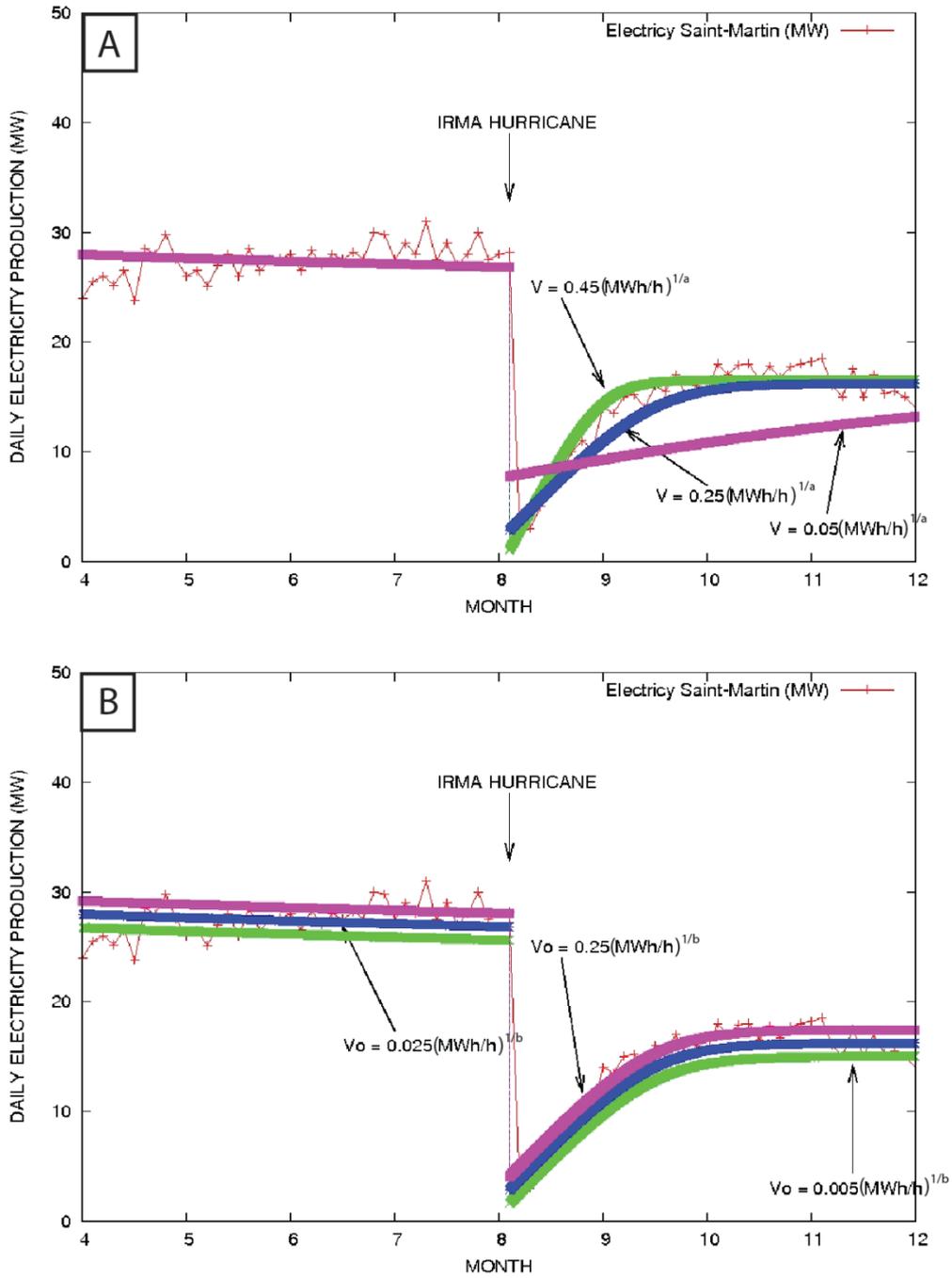

*Figure 7.* Influence of parameters V and V₀ on electric energy production. (A) Influence of parameter V on electric production in Saint-Martin, V₀ = 0.025

*(MWh/h)$^{1/b}$. (B) Influence of electric energy consumption $V_0$ on electric production, $V$ = 0.25 (MWh/h)$^{1/a}$. Other parameters in the simulation: $E_{max}$ = 56.6 MW, $\tau_0$ = 0.30, $\theta_t$ = 1 h, a = −0.054, b = −0.063, and L = 0.09 MWh.*

## 5. Discussion

*5.1. Hurricane destructive power and resilience of electric production*

In Puerto Rico, the influence of two hurricanes was observed in 2017 (Figure 4). The electric production was reduced by 30% of the initial electric capacity of Puerto Rico after Hurricane Irma (6 September) and it appeared to return "elastically" (i.e., linearly and rapidly) to the initial state (Figure 4). This was not the case following Hurricane Maria. The electric production was reduced by more than 90% in this second case.

The electric network and economic activities were not completely destroyed by the first hurricane. This could explain why two weeks after the first hurricane, the electric production was at the same level of 400 MW as the initial one (Figure 4). In the second case, the collapse was complete (i.e., the reduction in the electric production was above 90%) and the restoration was slower. The electric production was not impacted in the same way by the two hurricanes. The higher the lowering of the electric production during the hurricane, the longer the time required to reach the new equilibrium. The restoration rate of the electric production depends on the electric production capacity as well as the rate of electric production and consumption (see Equation 1 and Figures 5 to 7). The results suggest that the restoration of the electric production is not linear and that the restoration rate of the electric production depends on the amount of initial destruction. Furthermore, it depends on the adaptation and coping capacities as well (Medina et al., 2020).

After one year of Hurricane Maria, electric production of Puerto Rico attained the same values of approximately 400 MW (Figure 1). Nevertheless, the gross domestic product (GDP) of Puerto Rico was slightly lower in 2017 and 2018 than in 2016, suggesting that economic activity was not fully re-established to the level before Hurricane Maria [The World Bank, 2020], even if the electric production was almost completely restored (Figure 1).

*5.2. Electric production equilibrium*

A decrease of approximately 25% of the electric production was observed in Saint-Barthelemy and Puerto Rico, from the initial stable phase when $E = E_i$ until the new equilibrium phase after the collapse when $E = E_{is}$ (Figure 3 and 4; Table 1). In both cases, there was an approximate constant number of inhabitants before and after the hurricane. This reduction can be explained by a reduction in the activity of restaurants and shops, along with other social and individual activities (schools, administration, etc.). Buildings where these activities take place had been damaged (Rey et al., 2019).

In Saint-Martin, the decrease of 38% in electric production from $E = E_i$ to $E = E_{is}$ was higher than that in Puerto Rico and Saint-Barthelemy (Figure 2 and Table 1). The electrical network was almost completely restored after 1.5 months. Nevertheless, the restoration of electric production took longer (Figure 2), suggesting that other processes played a role. The remarkable reduction in electric production was also due to the reduction in economic activities. Unemployment increased by 18.5% in 2017 [IEDOM, 2019]. The number of hotels, guest houses, and apartments was limited in Saint-Martin because owing to their damage. There was a decrease of 25.9% in the number of tourists visiting the island from 2016 to 2017 [IEDOM, 2019]. By comparison, the number of tourists decreased by approximately 13% from 2016 to 2017 in Saint-Barthelemy [IEDOM, 2018]. The reconstruction of touristic infrastructures of Saint-Martin was not completed, to allow increases of tourism in 2018 and 2019. The tourism industry represents approximately 55% of the economic activity in Saint-Martin (IEDOM, 2019). Furthermore, Puerto Rico has a more diversified economy than Saint-Martin.

*5.3. Migration and electric production*

It has been estimated that approximately 7 000 to 8 000 inhabitants left Saint-Martin after the hurricane [Gustin, 2018; Desarthe et al., 2020]. The reduction of 20–23% of the children in public schools of Saint-Martin corresponds to the reduction of between 7 000 and 8 000 inhabitants [IEDOM, 2019; Desarthe et al., 2020]. The cause of this migration is mainly related to unemployment and difficulties of housing after the destruction caused by Irma.

Analogous to Saint-Barthelemy and Puerto Rico, where a decrease of 20–25% of the electric production was observed after the hurricane (Table 1), a decrease of approximately 20–25% of the electric production in Saint-Martin could be expected. Taking into account the global reduction of 38% of the electric production in Saint-Martin, it can be expected that the 13–18% of the electric production decrease is attributed to the migration of 7 000 to 8 000 inhabitants.

Historically, migration has been observed to occur in the Caribbean islands following environmental disasters (hurricanes, earthquakes) and economic or social crises (wars, dictatorships) [Audebert, 2003; Redon, 2006]. Haiti experienced various migration phases during the last decades [Amuedo-Dorantes et al., 2010]. Saint-Martin and Saint-Barthelemy experienced various migrations in or out of their territories during the last two centuries [Jeffry, 2010] as well. During the 1986–1990 period, there was an increase in the population of Saint-Martin from 8 000 to 28 000, which was mainly due to the influx from France, along with that from Haiti and the Dominican Republic after a new tax law was enacted in France [Seryans et al., 2017; Duvat, 2008; Pasquon et al., 2019].

The abrupt decrease in electric energy production is not specific to hurricanes and may also be caused by earthquakes or other natural hazards (floods). The intensity of the reduction depends on the amplitude of the catastrophe. The pattern of the electric production curves can be interpreted with Equations (1) and (2). The time between the catastrophe and the new equilibrium depends on (1) the time to rebuild the electric network, (2) the time to reconstruct the infrastructures that consume the electricity, (3) the time necessary for people who leave, to return, and (4) the time required to restore the economy. These times shown to be mixed in the electric production curves.

*5.4. Resilience of electric energy production*

If the undercapacity of electric production can limit the development of economic activity, this underdevelopment can slow down electric production restoration after major hurricanes. In Saint-Martin, the electric production capacity of the production units was re-established after approximately 25 days–1 month. Saint-Martin electric production has a smaller restoration rate $E_{is} / t_r$ than that of Saint-Barthelemy and Puerto Rico (Table 1). The slow rate of restoration of electric energy production was not limited by public electric infrastructure.

Additionally, the electric resilience ratio was higher in Saint-Barthelemy than in Saint-Martin (Table 1). This is in agreement with previous studies that observed that electric energy recovery rates were slower for low-income populations than for other groups (moderate-income to high-income) in Puerto Rico [Roman et al., 2019]. Indeed, Saint-Martin has a smaller GDP per capita than that of Saint-Barthelemy and Puerto Rico (Table 2). The correlation between wealth (GDP per capita) and resilience indicators (resilience ratio) suggests that the capacity of a territory to restore electric production is not independent of its wealth. This could depend on the difference in initial resistance of buildings and infrastructure, but also on the funding available for reconstruction (credit, insurances, solidarity funds, etc.) [Gargani, 2019; Gargani et al., 2020].

Nevertheless, the discrepancy in electric production per inhabitant between Saint-Martin and Saint-Barthelemy has not increased after the hurricane. Prior to the hurricane, there was an electric production of 1.65 kW/inhabitant in Saint-Barthelemy, higher than the 0.78 kW/inhabitant in Saint-Martin. Electric production per inhabitant is higher when the gross domestic product (GDP) is higher. The GDP is estimated to be 40 000 euros per capita in Saint-Barthelemy (IEDOM, 2018) and 16 600 euros per capita in Saint-Martin in 2014 [IEDOM, 2019] (Table 2). After Hurricane Irma, the electric production was of 1.2 kW/inhabitant in Saint-Barthelemy, whereas in Saint-Martin it was of 0.63 kW/inhabitant, considering a population of approximately 27 000 inhabitants (after the migration of 8 000 inhabitants). The electric production per capita was 2.1 times higher in Saint-Barthelemy than in Saint-Martin prior to the hurricane, and it was approximately 1.9 times higher after the hurricane. Evidently, the hurricane did not change the gap in electric production per capita between Saint-Barthelemy and Saint-Martin.

## 6. Conclusions

We investigated the evolution of electric energy production after major hurricanes occurred in 2017 in three Caribbean islands: Saint-Martin, Saint-Barthelemy, and Puerto Rico. Following these hurricanes, the electric energy production decreased abruptly. A resilience time of several months (1 month < $t_s$ < 5 months) was necessary to attain a new electric energy production equilibrium $E_{is}$ that was lower than the initial electric energy production $E_i$. The resilience time $t_r$ appeared to be proportional to the electric energy production capacities before the collapse, whereas the other resilience indicators were proportional to the wealth (GDP). The new stable electric production $E_{is}$ ranged between 25 and 38% less than the initial electric production. This reduction in electric production is mainly due to a decrease in the electric consumption per capita. The reduction in electric energy consumption per capita was 20–25%, suggesting a decrease in the social and economic activities of approximately 20–25% during the post-catastrophe equilibrium phase. Nevertheless, in the case of Saint-Martin, part of the reduction in the electric energy production was due to a reduction of the population owing to their migration outside the islands. The migration of approximately 8000 inhabitants from Saint-Martin reduced the electric energy production $E$ by 13–18%. The electric production can be used as an indicator of the resilience of social and economic activities.

**Declaration of interest**: none

Acknowledgement: This study has been funded by an ANR grant (French National Agency for Research, RELEV). The funding source had no involvement in the design and interpretation of the data in the writing of the report.